# How many interchanges does the selection sort make for iid geometric(p) input?

**Debasish Sahani[1] and Soubhik Chakraborty[2]**
**[1]Department of Mathematics, Indian School Mines University, Dhanbad, India**
**[2]Department of Applied Mathematics, BIT Mesra, Ranchi-835215, India**

**ABSTRACT:** The note derives an expression for the number of interchanges made by selection sort when the sorting elements are iid variates from geometric distribution. Empirical results reveal we can work with a simpler model compared to what is suggestive in theory. The morale is that statistical analysis of an algorithm's complexity has something to offer in its own right and should be therefore ventured not with a predetermined mindset to verify what we already know in theory. Herein also lies the concept of an empirical O, a novel although subjective bound estimate over a finite input range obtained by running computer experiments. For an arbitrary algorithm, where theoretical results could be tedious, this could be of greater use.

## 1 Introduction

Selection sort, also called replacement sort, is one of the simplest internal sorting algorithms, about 60% faster than bubble sort but inferior to insertion sort which is about twice faster than bubble sort. It is not efficient for sorting items higher than 1000. Its worst and average case complexity are both $O(n^2)$. It works by selecting the smallest unsorted item remaining in the list, and then swapping it with the item in the next position to be filled. For a comprehensive literature on sorting algorithms, see [Knu2000]. For sorting algorithms with reference to distribution theory, see [Mah00]. For a comparison between selection sort and bubble sort with emphasis on parameterized complexity, see [C+07]. The present note derives an expression for the number of interchanges made by selection sort for iid variates from geometric distribution. Empirical results reveal we can work with a simpler model compared to what is suggestive in theory. Although trivial, we first rule out continuous probability distribution input for the present study.

## 2. Continuous probability distribution input

Let a (1), a (2)…a (n) are independent and identically distributed random variables from some continuous distribution. Denoting P(.) as the probability function.

We have, due to continuity,

$$P\,[a\,(i) = a\,(j)] = 0 \tag{1}$$

Also:

$$P[a(i)>a(j)] = 1-P[a(i)<a(j)]-P[a(i)=a(j)] \tag{2}$$

Applying (2) in (1) yields

$$P\,[a\,(i)>a\,(j)] = 1-P\,[a\,(i)<a\,(j)] \tag{3}$$

Since

$$P\,[a\,(i)>a\,(j)] = P\,[a\,(i)<a\,(j)]$$

due to independence of a (i)'s.

So (3) can be rewritten as

$$2*P\,[a\,(i)>a\,(j)] = 1$$

Or, in other words, the probability of an interchange

$$P\,[a\,(i)>a\,(j)] = 1/2$$

which is independent of the parameter of the continuous distribution.

So the expected number of interchanges in selection sort is given by

$$n(n-1)/2*p[a(i)>a(j)] = n(n-1)/2*1/2 = n(n-1)/4$$

independent of the parameter of the input distribution. Here n(n-1)/2 is the number of comparisons.

## 3. Discrete probability distribution input

Let a (1), a (2)….a (n) are independent and identically distributed discrete random variables.

Due to independence of a(i) and a(j), we again have

$$P[a(i)>a(j)]=P[a(i)<a(j)] \quad (4)$$

Since (2) always holds, no matter what the nature of random variables might be, hence applying (4) in (2), we have

$$2P[a(i)>a(j)]=1-P[a(i)=a(j)] \quad (5)$$

Now

$P[a(i)=a(j)]= \Sigma P[a(i)=r, a(j)=r]$, summed over all values r can take

$$=\Sigma P[a(i)=r]P[a(j)=r]$$

due to independence of a(i) and a(j)

$$=\Sigma\{f(a)\}^2$$

where f(a) is the probability mass function of a.

When a(i)'s are independent and identically distributed (iid) geometric variates with parameter p, then the probability mass function of a(i)'s is given by

$$f(a)=p(1-p)^r,\ r=0,1,2,\ldots$$

In this case,

$$P[a(i)=a(j)] = \Sigma\{ p(1-p)^r \}^2, \quad (6)$$

the summation extending over r=0, 1, 2, 3…

If the a(i)'s are input for the selection sort algorithm, which has n(n-1)/2 comparisons, the expected number of interchanges

= number of comparisons x probability of interchange in a single comparison

$$= n(n-1)/2 \times P[a(i)>a(j)] = n(n-1)/2 \times ½[1 - \Sigma\{ p(1-p)^r \}^2 ],$$

the summation extending over r=0, 1, 2, 3, combining (5) and (6). This is the desired expression.

This expression involves both n and p and hence we infer that computational complexity must in cases such as this be expressed not only in terms of the input parameter characterizing the input size (n in our case) but also in terms of input parameter(s) characterizing the input probability distribution (p in our case).

## 4. Experimental Results

Can we write $C_{avg}(p)=O_{emp}(p^3)$ for a fixed n?

Following are some interesting empirical results on selection sort for geometric distribution input (simulated over100 trials). The random variable c measures the number of interchanges. We have calculated the mean, standard deviation and coefficient of variation of c abbreviated as meanc, sdc and cvc respectively. Their formulae are clear from the QBASIC code itself which we executed in our P4 system. We fixed n at 1000, as this is roughly the maximum value of n up to which this algorithm has been found to be efficient. For higher n, we might use insertion sort if simplicity is important or quick sort if speed is important. Our findings are summarized in Table 1 followed by the code.

**Table 1: Table showing the dependence of c on p for fixed n=1000**

| p | mean c | sd c | cv c |
|---|---|---|---|
| .1 | 30590.93 | 1785.8720 | .05838 |
| .2 | 17548.08 | 1294.4680 | .0737669 |
| .3 | 12175.57 | 1035.9940 | .0850879 |
| .4 | 9110.45 | 784.3701 | .0860956 |
| .5 | 6832.90 | 750.3445 | .1098135 |
| .6 | 5336.36 | 602.0993 | .1128296 |
| .7 | 4192.42 | 588.1761 | .1402951 |
| .8 | 3116.07 | 417.2080 | .1338892 |
| .9 | 2164.99 | 353.7879 | .1634132 |

**Interpretation**: It is evident that as p increases mean c decreases. This is because; an increase in p will imply that the number of failures is likely to be less preceding the first success. This means the observed range of values of the sorting items, which are iid geometric (p) variates, will be less resulting in more ties and consequently fewer interchanges. Although the exact dependence of mean c on p bears a non-polynomial relationship, as derived in the previous section, we did experiment with polynomial fits of degree 2, 3 and 4 (fig. 1-3 respectively). The logic behind our experiments centers on *Weierstrass'* theorem which states that any complex curve can be approximated by a polynomial of a suitable degree, under continuity assumption, within a given range. Here the argument p is continuous and lies within a fixed range [0, 1] but the dependent variable c is both discrete and random (see the preface of [Mah00]), so it is better to apply regression analysis than opt for an interpolating polynomial. Also, we wish to catch the general trend of the population rather than over-fit by forcing a polynomial to pass through all the points.

The following is the QBASIC code used in P4:

```
REM SELECTION SORT WITH GEOMETRIC VARIATE INPUT
CLS
n = 1000
INPUT "enter p"; p
DIM a(n)
s = 0: ss = 0
FOR trial = 1 TO 100
FOR i = 1 TO n
x = 0
  10 y = RND
IF y < p THEN a(i) = x: GOTO 20
x = x + 1
GOTO 10
  20 NEXT i
'selection sort begins'
c = 0
FOR i = 1 TO n - 1
FOR j = i + 1 TO n
IF a(i) > a(j) THEN SWAP a(i), a(j): c = c + 1
NEXT j
NEXT i
s = s + c: ss = ss + c * c
NEXT trial
meanc = s / 100
sdc = SQR(ss / 100 - meanc * meanc)
cvc = sdc / meanc
PRINT meanc, sdc, cvc
END
```

From fig 1-3, it is clear that the third degree polynomial is the simplest that is working well. Polynomials of degree four and higher are not recommended as we have to bear in mind the problem of ill-conditioning associated with higher degree polynomials. Under-fitting and over-fitting are both do be avoided in statistical modeling. See also the concluding section.

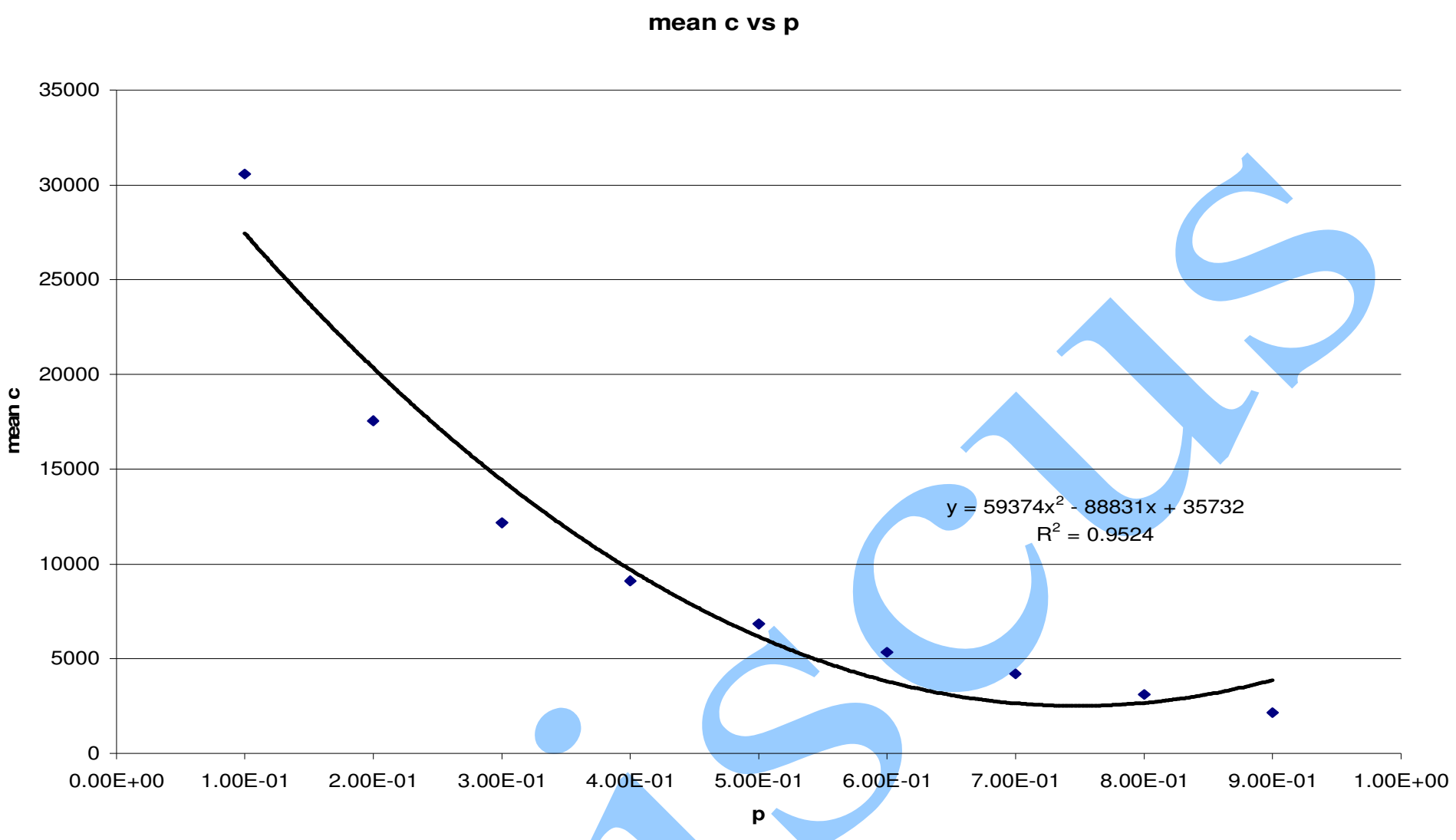

**Fig. 1: Experimenting with a polynomial of degree 2: underfitting**

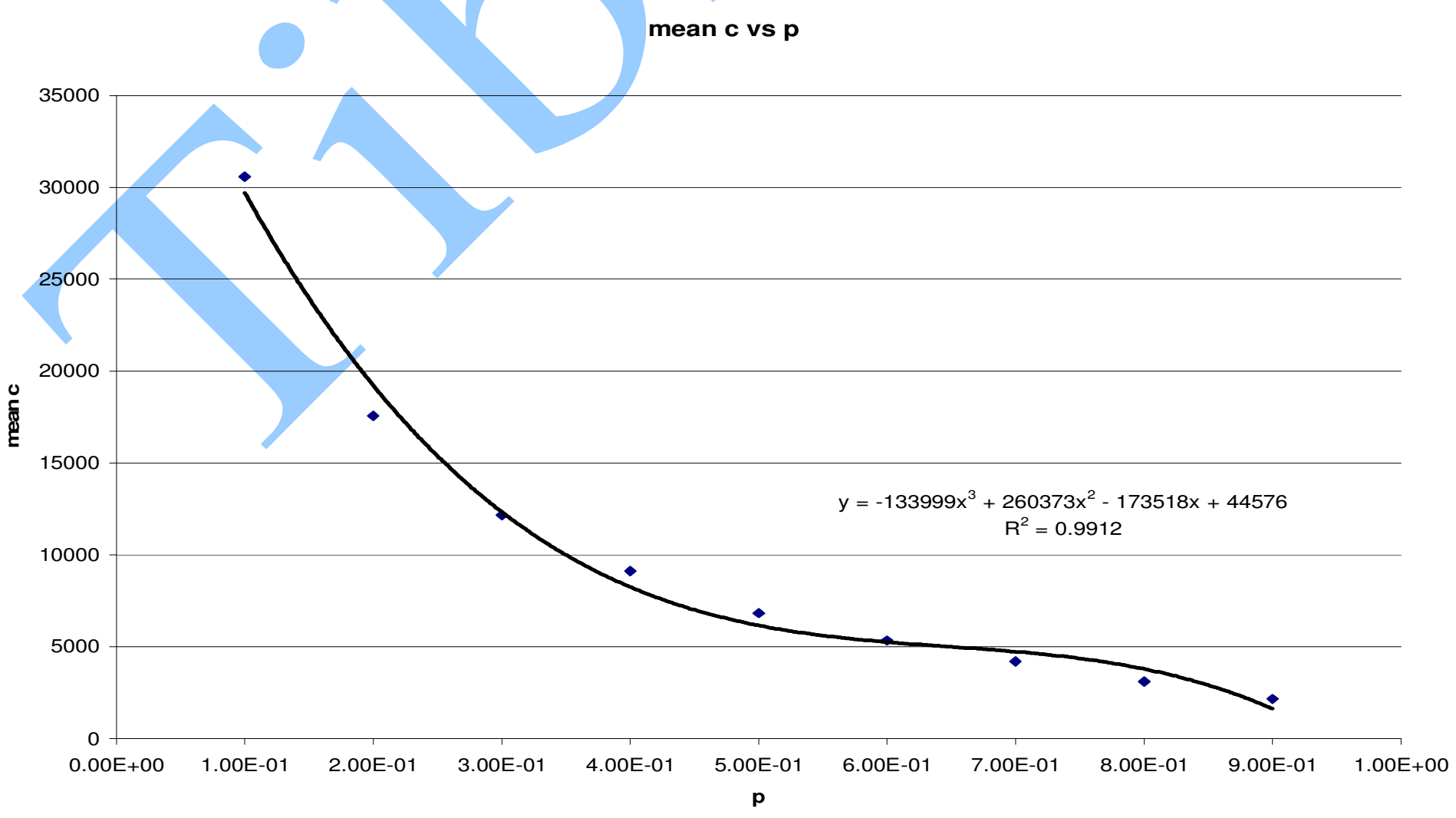

**Fig. 2: Experimenting with a polynomial of degree 3: simplest fit that works well**

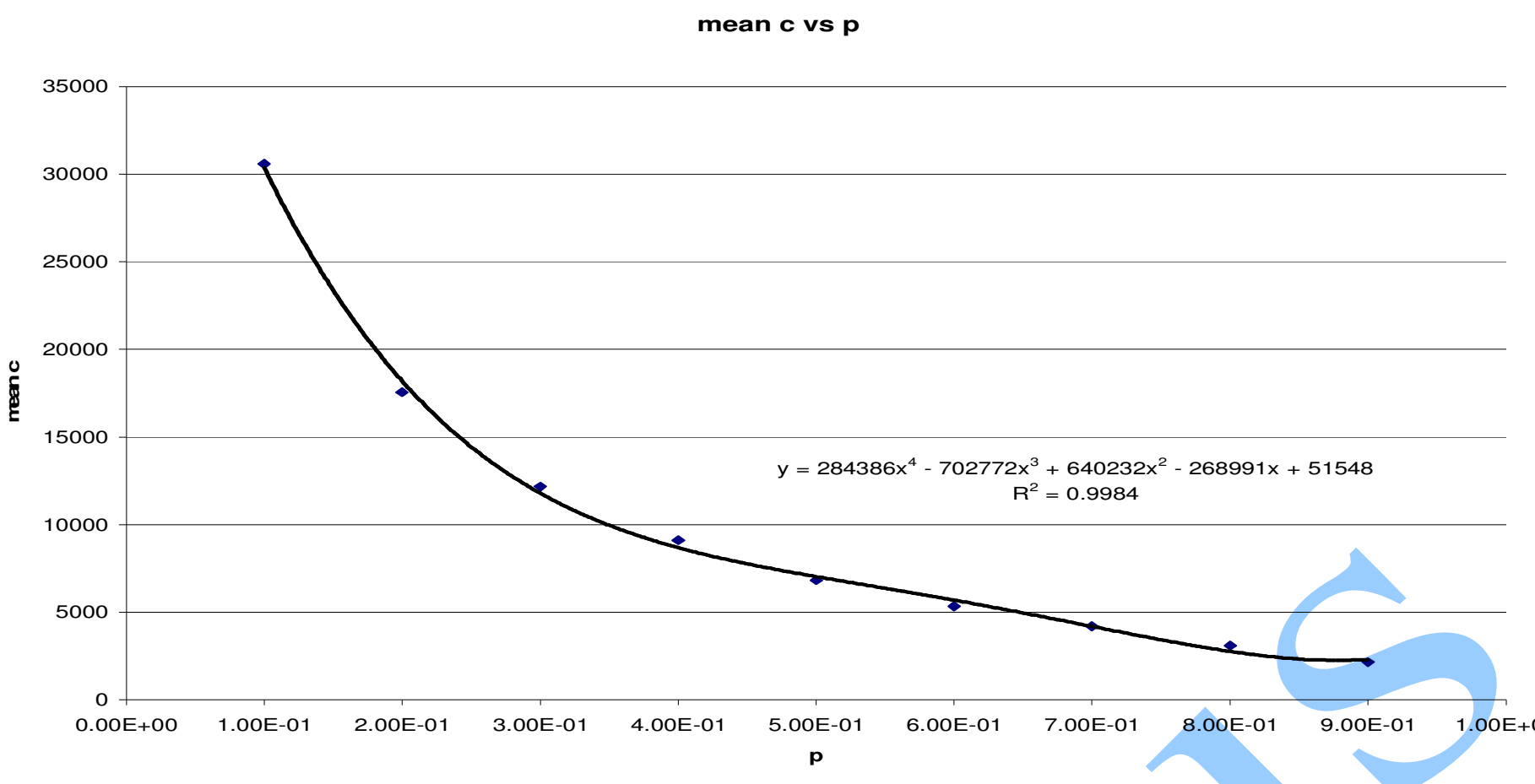

**Fig. 3: Experimenting with a polynomial of degree 4: overfitting?**

In the light of this finding, we re-experimented with the third degree polynomial fit using a trial version of SPSS package gathering some more information about the model. The reader is referred to tables 2-4 and fig. 4.

**Cubic fit (SPSS)**

**Table 2: Model summary (100$R^2$ gives the coefficient of determination, i.e., the % of variation in mean c accounted for by p through the fitted model)**

**Model Summary**

| R | R Square | Adjusted R Square | Std. Error of the Estimate |
|---|---|---|---|
| .996 | .991 | .986 | 1081.351 |

The independent variable is VAR00001.

**Table 3: ANOVA table applied to Regression Analysis**

**ANOVA**

| | Sum of Squares | df | Mean Square | F | Sig. |
|---|---|---|---|---|---|
| Regression | 6.548E8 | 3 | 2.183E8 | 186.660 | .000 |
| Residual | 5846595.079 | 5 | 1169319.016 | | |
| Total | 6.606E8 | 8 | | | |

The independent variable is VAR00001.

**Table 4: Standard error of the last square coefficients (VAR00001 is p)**
**Coefficients**

| | Unstandardized Coefficients | | Standardized Coefficients | t | Sig. |
|---|---|---|---|---|---|
| | B | Std. Error | Beta | | |
| VAR00001 | -173518.487 | 19171.152 | -5.229 | -9.051 | .000 |
| VAR00001 ** 2 | 260373.301 | 43399.090 | 8.046 | 6.000 | .002 |
| VAR00001 ** 3 | -133999.436 | 28639.647 | -3.769 | -4.679 | .005 |
| (Constant) | 44576.213 | 2337.970 | | 19.066 | .000 |

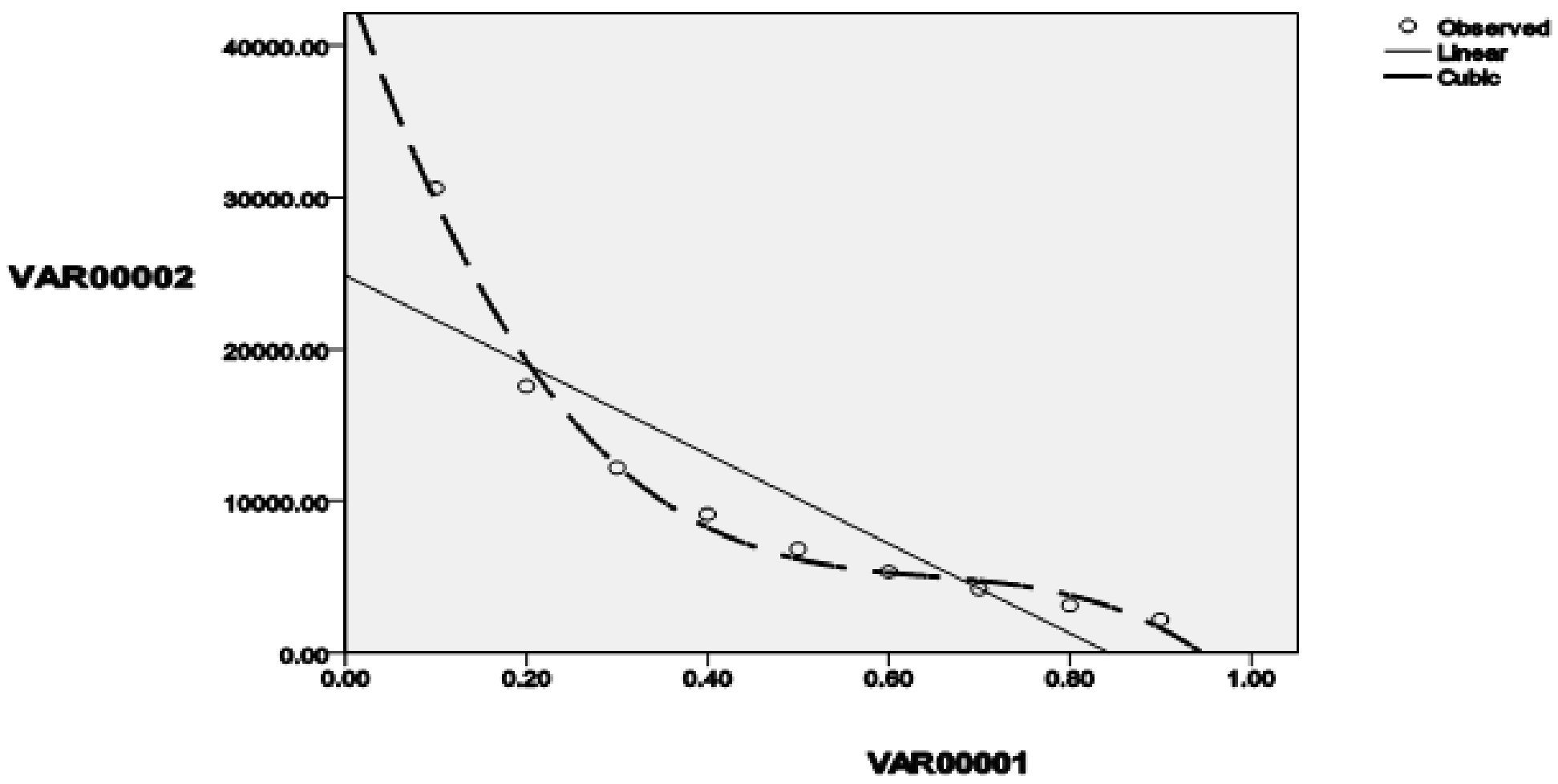

**Fig. 4: Cubic fit of mean c (Y-axis) against p (X-axis) using SPSS Var00002=mean c; var00001=p**

## 5. Concluding Remarks

The essence of curve fitting lies in catching the general trend exhibited by the observations and not in catching the observations themselves! It is clear from figure 4 that the third degree polynomial has captured the trend and that it is the simplest model to do so in our opinion. Since empirical O comes from a subjective empirical model fitted by statisticians to estimate complexity bounds over finite input ranges, given that in a computer experiment (a series of runs of a code for various inputs, Sacks et. al.

(1989)), the argument holds alike. While on one hand we stress the involvement of parameters of input distribution in complexity analysis of algorithms, we do also propose the estimation of these parameterized complexity bounds, which in general could be hard to derive or having expressions not very simplified or both, by empirical O. Using the concept of empirical O as defined in Chakraborty and Sourabh (2007) and noting carefully that in this case a count-based mathematical bound is being estimated over a finite range rather than a weight-based statistical bound (e.g. time can be a weight), we are permitted to write, subjectively, for a fixed n, $c_{avg}(p)=O_{emp}(p^3)$, the subscript emp indicating an empirical and subjective bound-estimate of parameterized complexity (here). The concept will be more useful for arbitrary algorithms, where theoretical analysis can be tedious.

A strong argument in favor of empirical O is that a bound that holds asymptotically may not hold over a finite range. Consequently, is it not rational for a programmer, who will be running computer experiments over finite input ranges surely, to select that algorithm for a given problem whose empirical O is stronger compared to that of another for the same problem, even if its theoretical bound is weaker? We hope our findings will lead to a fresh debate in algorithms. Moreover, let us also not forget that one of the major strengths of statistics lies in modeling. *All statistical models are subjective, but they arise from data that is objective--or nearly so--and often, at their center, stand some very elegant mathematical theorems--of course, free from bias.* The fundamental aspects of statistical modeling can been summed up very gently in three steps: first, fit a model to capture some phenomenon of the world around us; second, make an intelligent guess of the model parameters; and third, verify the goodness of the fit. We all know that statistics can be broadly divided into two categories: descriptive and inferential. In statistical modeling, both are involved--as we first describe a pattern (through modeling) and then infer about its validity (Klemens (2008); Chakraborty (2009)).

No statistical model is true or false, right or wrong. It is the statistician's motivation and expectation from the model and how far the model goes in fulfilling the same that matters in the final analysis. It was indeed our objective to convince the reader that such statistical analysis pertaining to algorithms and complexity as done here did not have a predetermined purpose of verifying what we already know in theory. The reader must appreciate that the statistical findings may have a new story to tell in its own right which cannot be thrown away. So the purpose achieved, we close the paper. [Concluded]

**Acknowledgement**

This paper was prepared as a short summer research project by the first author under the guidance of the second author. Both the authors thank Prof. P. K. Barhai, Vice Chancellor of BIT Mesra and Prof. N. C. Mahanti, Head, Department of Applied Mathematics, BIT Mesra.